\documentclass[aps,pre,reprint,groupedaddress]{revtex4-1}

\usepackage{graphicx}
\usepackage{lipsum}
\usepackage[modulo]{lineno}
\usepackage{amsmath}
\usepackage{amssymb}
\usepackage{hyperref}
\usepackage{xcolor}
\hypersetup{colorlinks=true, citecolor=blue}

\newcommand{\edit}[1] {#1}

\begin{document}


\title{\edit{Steady State Reduction} of generalized Lotka-Volterra systems in the
microbiome}


\author{Eric W. Jones}
\email[]{ewj@physics.ucsb.edu}
\author{Jean M. Carlson}
\affiliation{Department of Physics, University of California, Santa Barbara,
California 93106, USA}


\date{\today}

\begin{abstract}
    The generalized Lotka-Volterra (gLV) equations, a classic model from
    theoretical ecology, describe the population dynamics of a set of
    interacting species.  As the number of species in these systems grow in
    number, their dynamics become increasingly complex and intractable.  We
    introduce Steady State Reduction (SSR), a method that \edit{reduces} a gLV
    system of many ecological species into two-dimensional (2D) subsystems that
    each obey gLV dynamics and whose basis vectors are steady states of the
    high-dimensional model.  We apply this method to an experimentally-derived
    model of the gut microbiome in order to observe the transition between
    ``healthy'' and ``diseased'' microbial states.  Specifically, we use SSR to
    investigate how fecal microbiota transplantation, a promising clinical
    treatment for dysbiosis, can revert a diseased microbial state to health.
\end{abstract}


\maketitle



\section{Introduction}
The long-term behaviors of ecological models are proxies for the observable
outcomes of real-world systems.  Such models might try to predict whether a
pathogenic fungus will be driven to extinction \cite{BriggsVredenburg2010}, or
whether a microbiome will transition to a diseased state \cite{BucciXavier2014}.  In
this paper we explicitly account for this outcome-oriented perspective with
\textit{Steady State Reduction} (SSR). This method compresses a generalized
Lotka-Volterra (gLV) model of many interacting species into a reduced two-state gLV model
whose two unit species represent a pair of steady states of the original model.

This reduced gLV model is defined on the two-dimensional (2D) subspace spanned by
a pair of steady states of the original model, and the subspace itself is
embedded within the high-dimensional ecological phase space of the original gLV
model. We prove that the SSR-generated model is the best possible gLV
approximation of the original model on this 2D subspace.  The parameters of
the reduced model are weighted combinations of the parameters of the original
model, with weights that are related to the composition of the two
high-dimensional steady states. We note that SSR could be
extended to encompass three or more steady states, but the resulting
reduced systems would quickly become analytically opaque.
In Section \ref{sec:ssr} we describe SSR and
its implementation in detail. 

We apply this method to the microbiome, which consists of thousands of
microbial species in mammals \cite{RoundMazmanian2009}, and which exhibits distinct
``dysbiotic'' microbial states that are associated with diseases ranging from
inflammatory bowel disease to cancer \cite{Lloyd-PriceHuttenhower2016}.
Microbial dynamics are mediated by a complex network of biochemical
interactions (e.g. cellular metabolism or cell signaling) performed by
microbial and host cells \cite{WidderSoyer2016, PapinPalsson2004}.
Ecological models, including the gLV equations, seek
to consolidate these myriad biochemical mechanisms into nonspecific coefficients that
characterize the interactions between microbial populations.
We consider one
particular genus-level gLV model of antibiotic-induced \textit{C. difficile}
infection (CDI), which was fit with microbial abundance data from a mouse experiment
\cite{SteinXavier2013, BuffiePamer2012}. 

This CDI model exhibits steady states that correspond to
experimentally-observed outcomes of health (i.e.  resistance to CDI) or
dysbiosis (i.e. susceptibility to CDI). The transition between these healthy
and diseased states is difficult to effectively probe due to the high
dimensionality of the system, so previous analyses have been largely 
limited to numerical methods \cite{JonesCarlson2018}.  By reducing the
dimensionality of the original gLV model, SSR enables this transition to be
investigated with analytic dynamical systems tools. We demonstrate the fidelity
of SSR as applied to this CDI model in Section \ref{sec:stein}, and describe
the analytic tools accessible to reduced gLV systems in Section \ref{sec:2d}.

Finally, we use SSR to analyze the clinically-inspired scenario of
antibiotic-induced CDI.  Specifically, we examine the bacteriotherapy
\textit{fecal microbiota transplantation} (FMT), in which gut microbes from a
healthy donor are engrafted into an infected patient, and which has shown
remarkable success in treating recurrent CDI \cite{MerensteinLynch2014}. In
Section \ref{sec:2d} we implement FMT in the reduced model and successfully
revert a disease-prone state to health, and also find that the efficacy of FMT
depends on the timing of its administration. In Section \ref{11d_transient} we
show that this dependence on FMT timing, also present in the
experimentally-derived CDI model \cite{JonesCarlson2018}, is preserved under
SSR.

\section{Compression of generalized Lotka-Volterra systems} \label{sec:ssr}

\begin{figure}[b]
    \includegraphics[width=.4\textwidth]{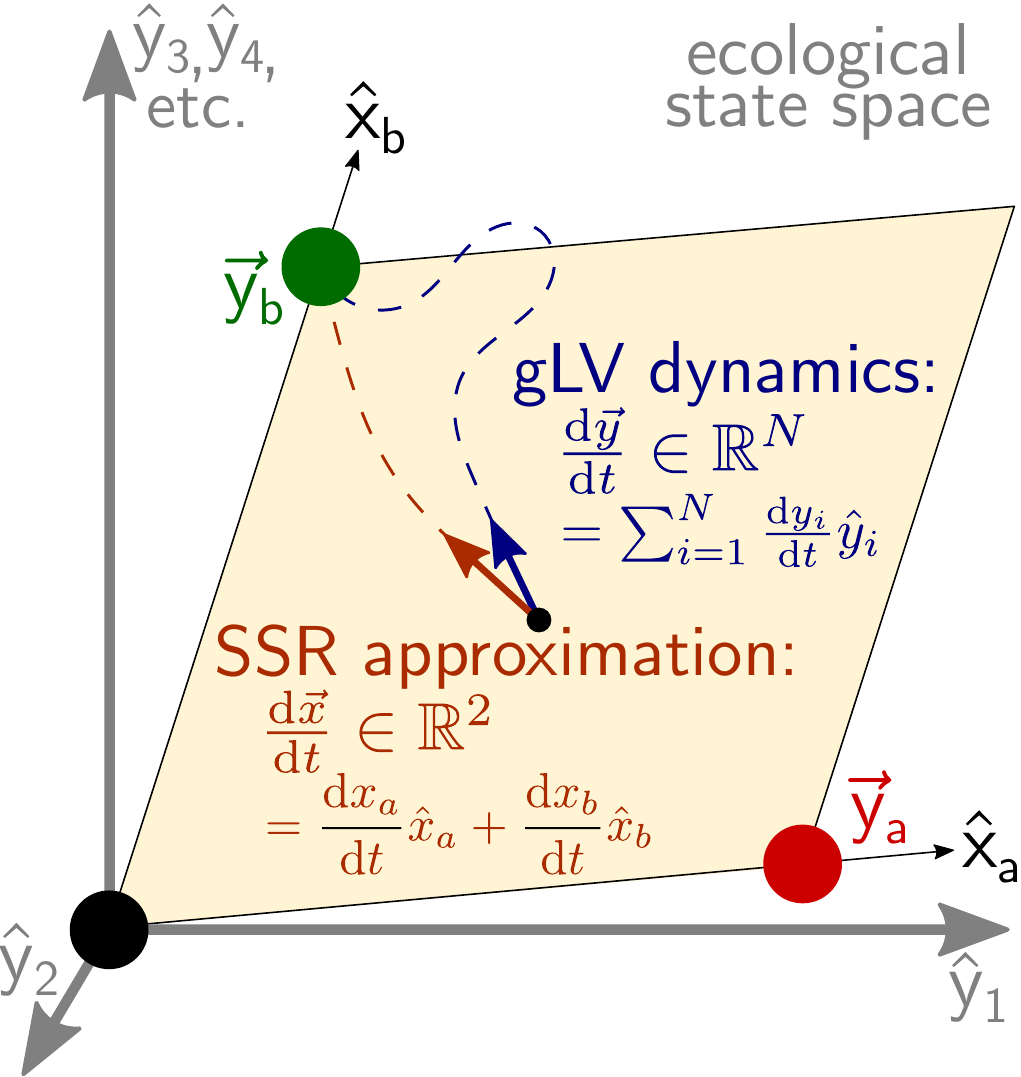}%
    \caption{Schematic of Steady State Reduction (SSR). A gLV system of $N$
    species (Eq.~(\ref{eq:gLV_N})) exhibits two steady states $\vec{y}_a$ and
    $\vec{y}_b$, characterized as diseased (red) and healthy (green).
    SSR identifies the
    two-dimensional (2D) gLV system defined on the
    2D subspace spanned by the two high-dimensional steady states (Eq.~(\ref{eq:gLV_2D})) 
    that
    best approximates the high-dimensional system. Specifically, 
    SSR prescribes 2D parameters (Eq.~(\ref{eq:SSR_params})) that minimize
    the deviation between the N-dimensional gLV dynamics
    $\text{d}\vec{y}/\text{d}t$ and the embedded 2D SSR-reduced
    dynamics $\text{d}\vec{x}/\text{d}t$.
    \label{fig:SSR}}
\end{figure}

We begin by demonstrating how to compress the high-dimensional ecological
dynamics of the generalized Lotka-Volterra (gLV) equations, given in
Eq.~(\ref{eq:gLV_N}), into an approximate two-dimensional (2D) subspace.
This process, called Steady State Reduction (SSR), is depicted schematically in
Fig.~\ref{fig:SSR}.  The idea behind SSR is to recast a pair of fixed points of
a high-dimensional gLV model as idealized ecological species in a 2D gLV model,
and to characterize the interactions between these two composite states by
taking a weighted average over the species interactions of the high-dimensional
system. Within this subspace, these reduced dynamics constitute
the best possible 2D gLV approximation of the high-dimensional gLV dynamics.

The gLV equations model the populations of $N$ interacting
ecological species $y_i$ as
\begin{equation}
    \frac{\text{d}}{\text{d}t} y_i(t) = y_i(t) \left( \rho_i + \sum_{j=1}^{N}
    K_{ij}y_j(t)  \right), \label{eq:gLV_N}
\end{equation}
for  $i \in 1,\ \ldots,\ N$. In vector form, these microbial dynamics are written 
$\frac{\text{d}\vec{y}}{\text{d}t} = \sum_{i=1}^N
\frac{\text{d}y_i}{\text{d}t} \hat{y}_i$.
Here, the growth rate of species $i$ is $\rho_i$,
and the effect of species $j$ on species $i$ is given by the interaction term
$K_{ij}$. In the following derivation, we assume this model observes
distinct stable fixed points $\vec{y}_a$ and $\vec{y}_b$.

Define variables $x_a$ and $x_b$ in the direction
of unit vectors $\hat{x}_a$ and $\hat{x}_b$ that parallel the two steady
states according to $\hat{x}_a \equiv
\vec{y}_a/\lVert \vec{y}_a \rVert_2$, and $\hat{x}_b
\equiv \vec{y}_b/\lVert \vec{y}_b \rVert_2$, where $\lVert \cdot \rVert_k$ is the
$k$-norm. The 2D gLV dynamics on the subspace spanned
by $\hat{x}_a$ and $\hat{x}_b$ are given by
\begin{align} \begin{split}
    \frac{\text{d}x_a}{\text{d}t} &= x_a (\mu_a + M_{aa}x_a + M_{ab} x_b),
    \quad \text{and} \\
    \frac{\text{d}x_b}{\text{d}t} &= x_b (\mu_b + M_{ba}x_a + M_{bb}x_b).
\label{eq:gLV_2D} \end{split} \end{align}
The \textit{in-plane dynamics} on this subspace in vector form
are defined to be
$\frac{\text{d}\vec{x}}{\text{d}t} = \frac{\text{d}x_a}{\text{d}t}\hat{x}_a +
\frac{\text{d}x_b}{\text{d}t}\hat{x}_b$.

SSR links the parameters of the in-plane dynamics to the high-dimensional gLV dynamics by
setting
\begin{equation}
    \begin{aligned}
        \mu_\gamma &= 
        \frac{ \vec\rho \cdot (\vec{y}_{\gamma}^{\ \circ 2} )}{\lVert \vec{y}_\gamma
        \rVert_2^2},
        \quad \ &&\text{ for } \ \gamma \in a, \ b, \quad \text{and} \\
        M_{\gamma \delta} &= 
        \frac{ (\vec{y}_{\gamma}^{\ \circ 2})^T K \vec{y}_\delta}{\lVert
        \vec{y}_\gamma
        \rVert_2^2 \lVert \vec{y}_\delta \rVert_2},
        && \text{ for } \ \gamma,\ \delta \in a, \ b.
    \end{aligned} \label{eq:SSR_params}
\end{equation}
Here, the Hadamard square represents the element-wise square of a
vector, defined as $\vec{v}^{\ \circ 2} = [v_1^2,\ v_2^2,\ \ldots,\ v_N^2]^T$.
The parameter definitions in Eq.~(\ref{eq:SSR_params}) are valid when $\vec{y}_a$
and $\vec{y}_b$ are orthogonal; when they are not, the cross-interaction terms
$M_{ab}$ and $M_{ba}$ become more complicated, and are given in Eqs.~(\ref{eq:c3})
and (\ref{eq:c6}) of the Appendix.

This choice of parameters minimizes the deviation between the in-plane and
high-dimensional
gLV dynamics $\epsilon = \lVert \frac{\text{d}\vec{y}}{\text{d}t} -
\frac{\text{d}\vec{x}}{\text{d}t} \rVert_2$ for any point on the subspace
spanned by $\hat{x}_a$ and $\hat{x}_b$. This is proved in the Appendix by showing
that, when evaluated with the SSR-prescribed parameter values of
Eq.~(\ref{eq:SSR_params}), $\frac{\partial \epsilon}{\partial c_i} = 0$ for
every coefficient $c_i \in \{\mu_a,\ \mu_b,\ M_{aa},\ M_{ab},\ M_{ba},\ M_{bb}
\}$, and that $\frac{\partial^2 \epsilon}{\partial c_i \partial c_j} > 0$ for
every pair of coefficients $c_i$ and $c_j$. 

Under this construction, the high-dimensional steady states $\vec{y}_a$ and
$\vec{y}_b$ have in-plane steady state counterparts at $(\lVert \vec{y}_a
\rVert_2,\ 0)$ and $(0,\ \lVert \vec{y}_b \rVert_2)$, respectively. It is for
this reason we call this method \textit{Steady State Reduction}. Further, if
$\vec{y}_a$ and $\vec{y}_b$ are stable and orthogonal, then the corresponding
2D steady states are stable as well, which guarantees the existence of a
separatrix in the reduced 2D system. These properties are shown in the
Supplementary Information \footnote{Supplementary
Information available at ****}, which includes many other calculations
that accompany the results of this paper. \edit{We provide a Python module that implements SSR on arbitrary
high-dimensional gLV systems in the Supplementary Code
\footnote{Supplementary
Code used to implement SSR and generate Fig.~2 available at
\texttt{github.com/erijones/ssr\_module}.}.}

If the ecological dynamics of the system lie entirely on the plane spanned by
$\hat{x}_a$ and $\hat{x}_b$, the SSR approximation is exact. In this case,
the plane contains a slow manifold on which the ecological dynamics evolve.
Therefore, the dynamics generated by SSR result from a
linear approximation of the slow manifold.

\section{Steady state reduction applied to a microbiome model}
\label{sec:stein}

Thousands of
microbial species populate the gut microbiome \cite{RoundMazmanian2009}, but
for modeling purposes it is common to coarse-grain at the genus or phylum
level. Recently, many experimentally derived gLV microbiome models have been
constructed with tools such as MDSINE, a computational pipeline that infers gLV
parameters from time-series microbial abundance data \cite{BucciGerber2016}.
SSR is applicable to any of these gLV systems, so long as it exhibits two or
more stable steady states. 

We consider one such experimentally derived gLV model, constructed by
Stein et al.  \cite{SteinXavier2013}, that studies CDI in the mouse
gut microbiome. This model takes the same form as
Eq.~(\ref{eq:gLV_N}) and tracks the abundances of 10 different microbial genera
and the pathogen \textit{C.  difficile} (CD), all of which can inhabit the
mouse gut. The 11-dimensional (11D) parameters of this model were fit with data
from an experimental mouse study \cite{BuffiePamer2012}. The
parameters of this model, along with a sample microbial trajectory, are
provided in the Supplementary Information \cite{Note1}.

Despite the fact that this model did not resolve individual bacterial species,
it still captured the clinically- and experimentally-observed phenomenon of
antibiotic-induced CDI, suggesting that the true microbiome's dimensionality
could be approximated by an 11-dimensional model.  SSR further
simplifies the dimensionality of the microbiome: instead of
thousands of microbial species or even eleven dominant genera,
with SSR steady states of the microbiome (each of which are multi-species equilibrium
populations) are idealized as individual ecological populations. 

This CDI model exhibits five steady states that are reachable from
experimentally measured initial conditions \cite{SteinXavier2013}. In previous
work,
we identified which of these steady states were susceptible or resilient to
invasion by \textit{C. difficile} (CD) \cite{JonesCarlson2018}.  Based on this
classification, we interpret a CD-susceptible steady state $\vec y_a$ of
the 11D model as ``diseased,'' and interpret a CD-resilient steady state $\vec
y_b$ as ``healthy.'' Explicit details about each of these states are provided in
the Supplementary Information
\cite{Note1}. These two states are used to demonstrate SSR.

\begin{figure}[t]
    \includegraphics[width=.45\textwidth]{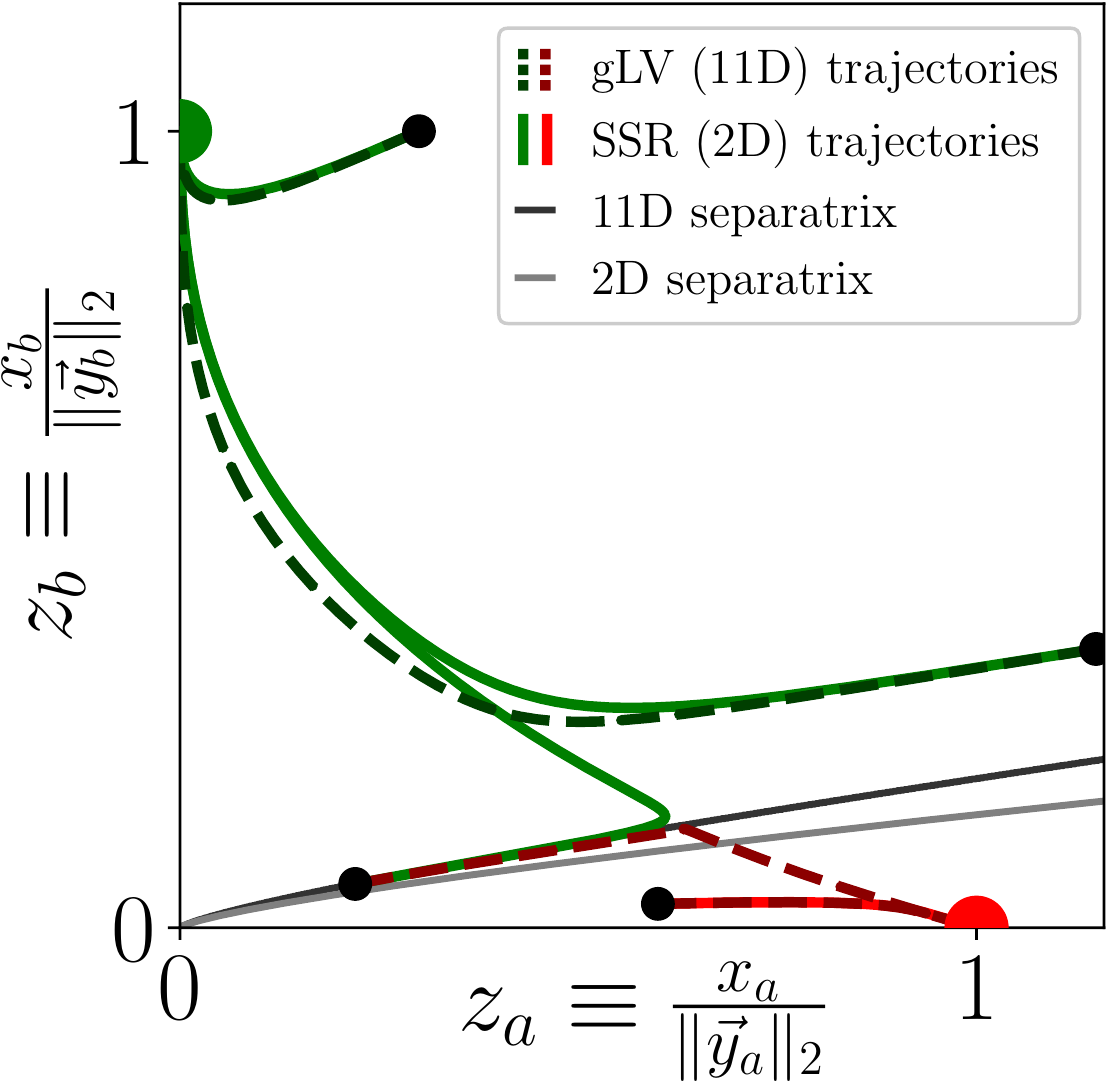}%
    \caption{Fidelity of Steady State Reduction (SSR). SSR is applied to an
    experimentally-derived 11-dimensional (11D) gLV model of \textit{C.
    difficile} infection (CDI) \cite{SteinXavier2013}. This model exhibits
    steady states $\vec{y}_a$ and $\vec{y}_b$ that are vulnerable (diseased,
    red) and resilient (healthy, green) to invasion by the pathogen \textit{C.
    difficile} \cite{Note1}. We consider 11D microbial trajectories whose
    initial conditions lie the plane spanned by these two steady states, and
    plot the in-plane projection of these trajectories (dashed lines).  The 2D
    SSR-generated dynamics (solid lines) are plotted alongside these
    high-dimensional trajectories.  The separatrix of each system is also
    plotted: as a proxy for the 11D separatrix (actually a 10-dimensional
    surface), the in-plane separatix (dark grey) is numerically generated and
    plotted; the 2D separatrix is exact and given in Eq.~(\ref{eq:sep}) (light
    grey). \edit{The code used to generate this figure is available in the
    Supplemental Code \cite{Note2}.}
    \label{fig4}}
\end{figure}

The reduced 2D parameters are generated according to Eq.~(\ref{eq:SSR_params}).
We introduce new scaled variables, $z_a = x_a / \lVert \vec{y}_a \rVert_2$, and
$z_b = x_b / \lVert \vec{y}_b \rVert_2$, so that the 2D system exhibits steady
states at $(1, \ 0)$ and $(0, \ 1)$. In Fig.~\ref{fig4}, trajectories of the
reduced system (solid lines) that originate from four initial conditions and
tend toward either the healthy (green) or diseased (red) steady states are
plotted. The 2D separatrix is also plotted (light grey), which divides the
basins of attraction of the two steady states, and which is derived in
Eq.~(\ref{eq:sep}) of the Section \ref{sec:2d}.

To compare the original and reduced models, we consider 11D trajectories that
originate from the 11D embedding of the four 2D initial conditions \cite{Note1}.  The
projections of these 11D trajectories onto the 2D subspace spanned by
$\vec{y}_a$ and $\vec{y}_b$ (dashed lines) are shown alongside the
corresponding 2D trajectories in Fig.~\ref{fig4}.  The in-plane 11D separatrix
is also shown (dark grey), which is numerically constructed by tracking the
steady state outcomes of a grid of initial conditions on the plane.

\edit{We note that $\vec{y}_a$ and $\vec{y}_b$ are nearly orthogonal. However,
in the Supplementary Information we demonstrate that the high-dimensional and
SSR-reduced trajectories and basins of attraction agree for four different
implementations of SSR; in two of these implementations  the pairs of steady
states were orthogonal, and in the other two they were not \cite{Note1}.  It is
important to understand when SSR is a good approximation, and under what
conditions it may be successfully applied--- this issue will be addressed in a
future publication (in progress). 
}

\edit{In the five realizations of SSR explored in this paper and in the
supplement,}
the basins of attraction and microbial trajectories are largely preserved
through SSR. Since the 11D system has been compressed (from 132 parameters to
6), it is not surprising that the low- and projected high-dimensional
trajectories do not exactly match. Even so, the basins of attraction agree
almost entirely, and the dynamical trajectories share clear similarities.  The
deviation between the original and reduced systems is examined in more detail
in the Supplementary Information \cite{Note1}.  The close agreement between the
original and reduced systems intimates the reductive potential of SSR. 

\section{Analysis of the 2D gLV equations} \label{sec:2d}
Having demonstrated a method of linking a high-dimensional gLV system to a 2D
gLV system via SSR, we now take advantage of the analytic accessibility of such
2D systems.
We consider
biologically relevant systems that exhibit competitive
dynamics by assuming $\mu_\alpha > 0$ for $\alpha \in a,\ b$, and $M_{\alpha
\beta} < 0$ for $\alpha,\ \beta \in a,\ b$.  These systems
exhibit two stable and homogeneous fixed points at $(-\mu_a/M_{aa},\ 0)$
and $(0,\ -\mu_b/M_{bb})$. In this case, the system will also possess a
hyperbolic fixed point at $(x_a^*,\ x_b^*)$ with $x_a > 0$ and $x_b > 0$, which
topologically guarantees the existence of a separatrix. 

In Section \ref{explicit} this separatrix is explicitly calculated for the
2D gLV system Eq.~(\ref{eq:gLV_2D}). This result, in conjunction with
SSR, allows for an efficient approximation of the high-dimensional separatrix.
Then, Section \ref{section_timing} explores the steady state and transient
dynamics of a nondimensionalized form of the 2D gLV system with
clinically-inspired modifications.

\subsection{Explicit form of the separatrix} \label{explicit}

The long-term dynamics of this system are dictated by the basins of attraction
of the stable steady states, and these basins are delineated by a separatrix
that, for topological reasons, must be the stable manifold of the hyperbolic
fixed point $(x_a^*, \ x_b^*)$. In Fig.~\ref{fig1} these basins are depicted 
topographically via isoclines of the split Lyapunov function $V(x_a, x_b)$
(lightly shaded contours), which acts as a potential energy landscape
\cite{Note1, HouStamova2013}.

The separatrix $h(x_a)$ may be analytically computed in a power series
expansion about the hyperbolic fixed point $(x_a^*,\ x_b^*)$,
\begin{equation}
    h(x_a) = \sum_{n=0}^\infty \frac{c_n}{n!}(x_a - x_a^*)^n, \label{eq:sep}
\end{equation}
which as an invariant manifold must satisfy \cite{Wiggins2003}
\begin{equation}
    \frac{\text{d} h(x_a)}{\text{d} x_a} = 
    \frac{\text{d} x_b}{\text{d} t} \bigg / \frac{\text{d} x_a}{\text{d} t},
    \label{eq:full_expansion}
\end{equation}
resulting in the recursive relations
\begin{equation}
\begin{split}
    c_0 &= x_b^*, \\
    c_{1} &= \frac{1}{2 M_{ab}x_a^*} \left[
        M_{bb}x_b^* - M_{aa}x_a^* \vphantom{\sqrt{1^2}}\right. \\
    &\quad \quad \quad \left. - \sqrt{(M_{bb}x_b^* - M_{aa}x_a^*)^2 + 4
    M_{ab}M_{ba}x_a^* x_b^*} \right], \\
    c_2 &= 
    \frac{2 c_1(M_{ba} + M_{bb} c_1 - M_{aa} - M_{ab} c_1)}{2x_a^*M_{aa} +
    3x_a^*M_{ab} c_1 - M_{bb} x_b^*}, \quad \text{ and} \\
    c_n &=
    \frac{1}{\left( n x_a^* M_{aa} + (n+1) x_a^* M_{ab} c_1 - M_{bb}
    x_b^* \right) } \\
    & \quad \times \left\{ n c_{n-1}(M_{ba} + M_{bb} c_1 -
    (n-1)(M_{aa} + M_{ab} c_1)) \vphantom{\sum_i^1} \right. \\
    & \quad +
    n! \sum_{\ell=2}^{n-1}\left[\frac{c_\ell}{\ell ! \ (n - \ell)!} \left(M_{bb}
    c_{n-\ell}- (n-\ell) M_{ab} c_{n - \ell}  \right. \right. \\
    & \quad \quad \quad \quad \quad \quad \left. \left. \left. - x_a^* M_{ab} c_{n - \ell +
    1}\right) \vphantom{\frac{1}{2}} \right] \vphantom{\sum_1^2} \right\},
    \quad \text{ for } n>2, \\
    \label{eq:sep}
\end{split}
\end{equation}
as derived in Eqs.~(S27-S38) \cite{Note1}.  This calculation allows
the \textit{a priori} classification of the fate of a given initial condition, without need
for simulation. We note that this algebraic calculation of the separatrix is
considerably faster than numerical methods that rely on relatively costly
quadrature computations.  Further, in conjunction with SSR, this analytic form
offers an efficient approximation to the in-plane separatrix of
high-dimensional systems.

\subsection{Dynamical landscape of the 2D gLV equations}
\label{section_timing}

\begin{figure}[t]
    \includegraphics[width=.4\textwidth]{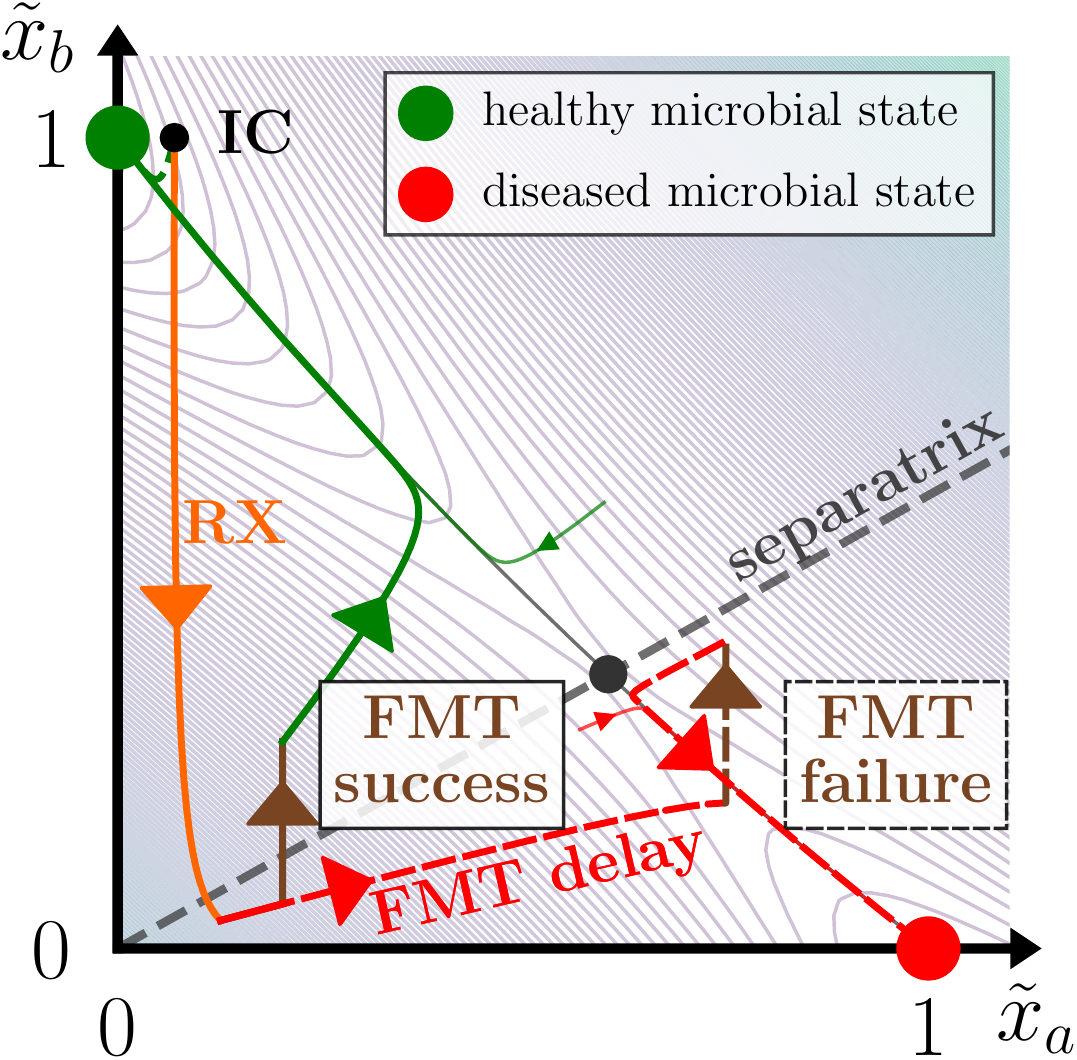}%
    \caption{The success or failure of fecal microbiota transplantation (FMT)
    depends on the timing of its administration in a two-state gLV system
    (Eq.~(\ref{eq:gLV_nondim})). We consider a clinically-inspired scenario that
    parallels antibiotic-induced CDI.
    First, a health-prone initial condition (IC) is depleted by antibiotics
    (RX, orange). If FMT (brown) is administered shortly after the antibiotics,
    the treatment steers the composition to a healthy state (FMT success). If
    FMT administration is delayed, the microbial trajectory instead attains the
    diseased state (FMT failure).  The basins of attraction of the healthy and
    diseased steady states are delineated by the separatrix Eq.~\ref{eq:sep}, and isoclines of
    the potential energy landscape (light contours) are given by the
    split Lyapunov function Eq.~(S47) \cite{HouStamova2013, Note1}.
    \label{fig1}}
\end{figure}

Next, we analyze a two-state implementation of the gLV equations that parallels the
clinically-inspired scenario of antibiotic-induced CDI. In this scenario,
antibiotics deplete a health-prone initial condition, requiring administration
of FMT in order to recover, as in Fig.~\ref{fig1}. FMT is implemented in the 2D
gLV model by adding a transplant of size $s$ composed of the healthy steady state $(0,\
1)$ to an evolving microbial state at a specified time following administration
of antibiotics.

We consider a nondimensionalized form of the gLV equations
Eq.~(\ref{eq:gLV_2D}) and designate nondimensionalized variables with a tilde
\cite{Note1}. Therapeutic interventions of
antibiotics and FMT are included in this model in a manner consistent with previous
approaches \cite{SteinXavier2013, JonesCarlson2018}. 
In all, this
clinically-inspired two-state gLV model is given by
\begin{align} \begin{split}
    \frac{\text{d}\tilde{x}_a}{\text{d}t} &= \tilde{x}_a (1 - \tilde{x}_a -
    \tilde M_{ab} \tilde{x}_b) \\
    &\quad + \tilde{x}_a \varepsilon_a u(t) + w_a \delta(t - t^*), \ \text{
        and}\\
    \frac{\text{d}\tilde{x}_b}{\text{d}t} &= \tilde{x}_b (\mu_b - \tilde
    M_{ba}\tilde{x}_a - \tilde{x}_b) \\
    &\quad + \tilde{x}_b \varepsilon_b u(t) + w_b \delta(t - t^*),
\label{eq:gLV_nondim}\end{split} \end{align}
which includes optional antibiotic administration $u(t)$ operating with
efficacy $\vec{\varepsilon}$, and optional FMT with transplant $\vec{w}$
administered instantaneously at time $t^*$.  

In the absence of antibiotics and FMT, the dynamical system
Eq.~(\ref{eq:gLV_nondim}) exhibits three nontrivial steady states at $(1, \
0)$, $(0, \ \mu_b)$, and $(\tilde x_a^*, \ \tilde x_b^*)\equiv(\frac{1 -
\tilde{M}_{ab} \tilde{\mu}_b}{1 - \tilde{M}_{ab}\tilde{M}_{ba}},\
\frac{\tilde{\mu}_b - \tilde{M}_{ba}}{1
- \tilde{M}_{ab}\tilde{M}_{ba}})$. To simplify the presentation of our
results in the main text we assume $\mu_b = 1$, though 
this assumption is relaxed in the Supplementary Information \cite{Note1}.

\begin{figure}[t]
    \includegraphics[width=.47\textwidth]{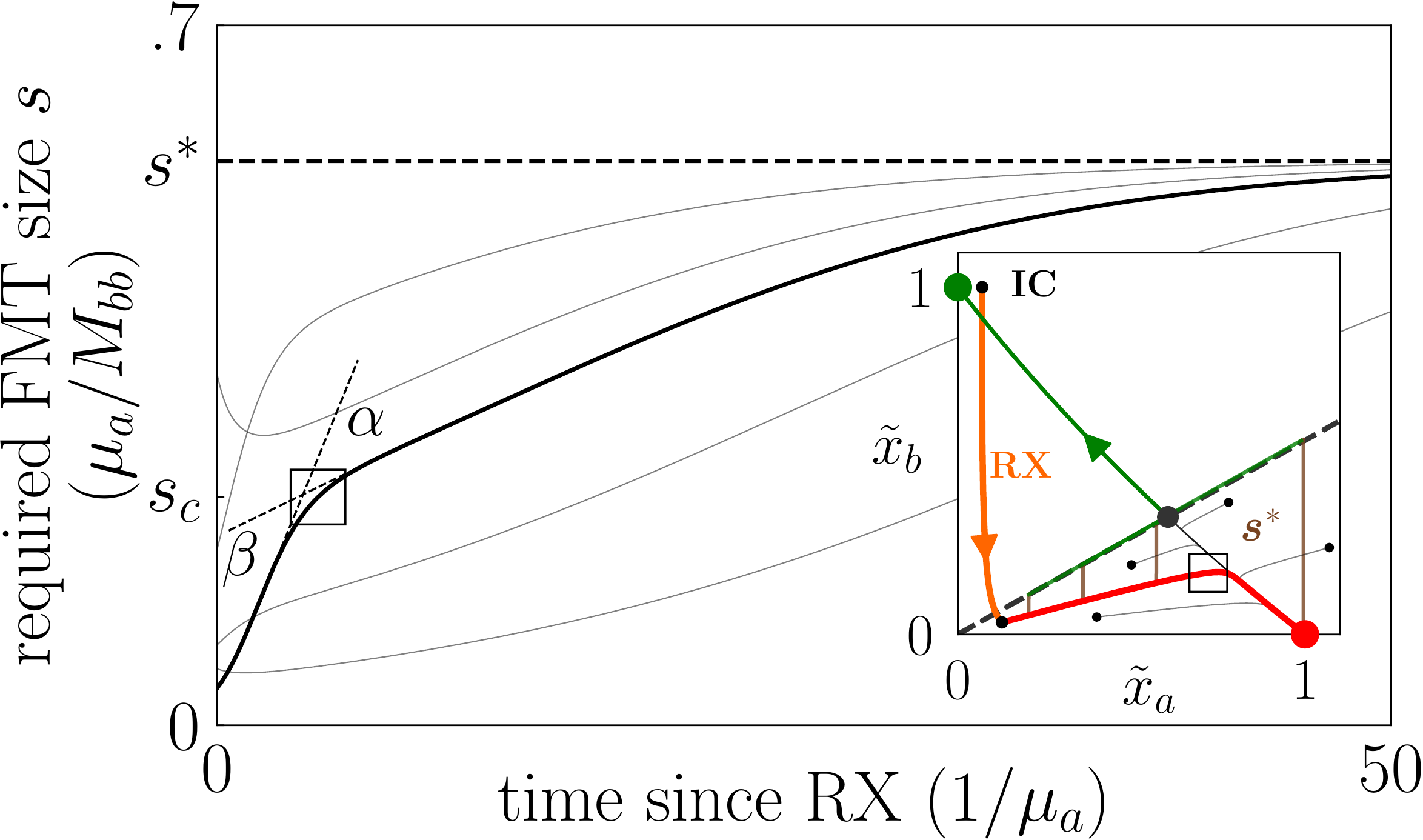}%
    \caption{The FMT transplant size needed to revert an antibiotic-depleted
    state back to health grows as FMT administration is delayed.
    The minimum FMT transplant size required to cure five distinct disease-prone
    microbial trajectories,
    each evolving according to Eq.~(\ref{eq:gLV_nondim}), are calculated and
    plotted. As
    trajectories attain the diseased steady state, the required transplant size
    approaches $s^*$. The required transplant size changes at two different
    rates, $\alpha$ and $\beta$, with the crossover point between these two
    rates at size $s_c$ indicated by a hollow square.  The transplant size dynamics
    $\text{d}s/\text{d}t$ as well as the rates $\alpha$ and $\beta$ are derived
    in Eq.~(\ref{eq:dudt}) and the surrounding text.
    \label{fig5}}
\end{figure}

\begin{figure}[t]
    \includegraphics[width=.47\textwidth]{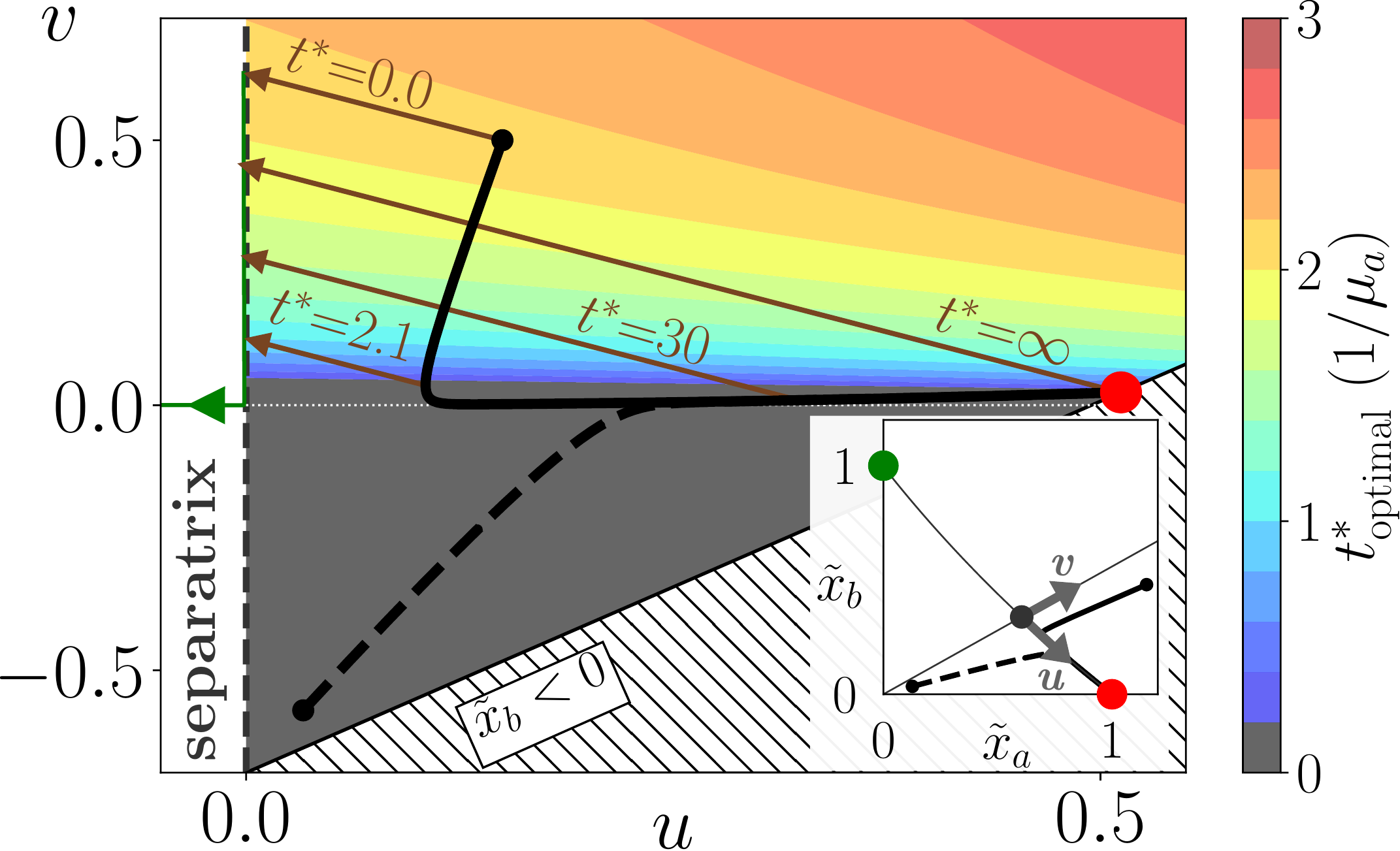}%
    \caption{The role of timing in FMT administration. For antibiotic-depleted
    disease-prone
    initial conditions in which antibiotics have been cleared ($u(t) = 0$),
    FMT is most effective when administered immediately
    ($t^*_{\text{opt}} = 0$, grey) or nearly immediately ($t^*_{\text{opt}} > 0$, colored)
    following antibiotic administration. 
    The
    optimal transplant time $t^*_{\text{opt}}$ is computed for any initial condition ($u_0, \
    v_0$) (colorbar) according to Eq.~(S88) of \cite{Note1}, which can reduce
    to Eq.~(\ref{eq:tstar}).  Two representative microbial trajectories are
    plotted in $(u, \ v)$ (main panel) and $(x_a, \ x_b)$ (inset) coordinates.
    For $v_0 > 0$ four possible FMT transplants are shown, including the optimal
    one that occurs at $t^*_{\text{opt}} = 2.1$.  For $v_0 < 0$ it is always best to
    administer FMT immediately following antibiotic administration.
    \label{fig3}}
\end{figure}

As before, suppose the variable $\tilde x_a$
corresponds to a diseased state, and $\tilde x_b$ corresponds to a healthy state.
Also assume the
transplant $\vec{w}$ consists of exclusively healthy microbes so that $w_a = 0$.
Figs.~\ref{fig1}, \ref{fig5}, and \ref{fig3} are generated with
parameter values $\tilde{M}_{ab} = 1.167$ and $\tilde{M}_{ba} =
1.093$, which give typical results.

Altering the fate of an initial condition requires crossing the separatrix by
some external means, which is achieved through FMT. Fig.~\ref{fig1} shows two
microbial time courses in which long-term outcomes are determined by the timing
of FMT administration.  At each point along a microbial trajectory in the
diseased basin of attraction, the minimum FMT size $s$ required to transfer the
microbial state into the healthy basin of attraction is calculated.  We use
this metric to quantify our notions of ``FMT efficacy.'' 

In clinical practice FMT administration varies in transplant size, transplant
composition, and how many transplants are performed. Further, it is unclear how
these factors influence the success of FMT \cite{HotaPoutanen2018}.  For the
purposes of this paper, we consider a hypothetical FMT treatment of size $s_t$
(i.e. a horizontal cut across Fig.~\ref{fig5}) and describe how its success
depends on the timing of its administration. 

Fig.~\ref{fig5} presents the minimum FMT size $s$ as a function of time (main
panel) for several trajectories that originate in the diseased basin of
attraction (inset), including the main trajectory of Fig.~\ref{fig1}.  This
minimum required FMT size increases with time--- often dramatically--- and
there are two discernible rates of increase, denoted $\alpha$ and $\beta$ in
Fig.~\ref{fig5}.  These two rates are related to the fast and slow manifolds of
the ecological system, which in turn govern the minimum required transplant
size dynamics over time.  

To reflect the importance of the separatrix in dictating the microbial
dynamics, we change coordinates to the eigenvectors $(u,\ v)$ of the hyperbolic
steady state, shown in Fig.~\ref{fig3} (inset). In these coordinates the
$v$-axis corresponds to the separatrix, and $u$ is proportional to the
minimum FMT size required for a successful transplant $s$, such that $s = u
/(\hat{u} \cdot \hat{x}_b)$, where $(\hat{u}, \ \hat{v})$ and $(\hat{x}_a, \
\hat{x}_b)$ are the unit vectors associated with their associated coordinates.

In this new $(u, v)$ basis, the 2D gLV equations become
\begin{align} \begin{split}
    \frac{\text{d}u}{\text{d}t} &= A_{10} u - A_{11} u v - A_{20} u^2, \quad
    \text{and}  \\
    \frac{\text{d}v}{\text{d}t} &= -B_{01} v - B_{02} v^2 + B_{20} u^2,
\end{split} \label{eq:gLV_uv} \end{align}
where each coefficient is a positive algebraic function of the original gLV
parameters given analytically in Eqs.~(S60-S74) \cite{Note1}.
When $\mu_b \neq 1$, these equations contain additional quadratic terms
described in \cite{Note1} that account for the nonlinearity of the separatrix.
In the small $u$ and small $v$ limit this model reduces to the
linearization about the hyperbolic fixed point.  Near this fixed point there is
a separation of time scales between $u$ and $v$ ($B_{01} / A_{10} > 1$ always, with
median of 5.9 and IQR of [2.7, 9.1] over random parameter draws \cite{Note1}),
indicating that there are inherent fast and slow manifolds in this system.

This coordinate change also reveals the role of timing in FMT administration,
since the minimum required transplant size $s$ is precisely governed by
Eq.~(\ref{eq:gLV_uv}), by proxy of $u$. To demonstrate this analytically, we
consider an initial condition condition $(u_0, \ v_0)$ that is located near the
fast manifold in a system with clear separation of time scales, so that (i)
$B_{20} u_0^2$ is negligible, (ii) $A_{10} << B_{01} $, and (iii) $B_{02} v_0^2 <<
B_{01} v_0$ (though this assumption is relaxed in Eq.~(S87) \cite{Note1}). In
this case, the dynamics in the fast $\hat{v}$ direction are approximately
$v(t) \approx v_0 \text{e}^{-B_{01} t}$, and the required transplant size
dynamics reduce to 
\begin{equation}
    \frac{\text{d}s}{\text{d}t} = s \left( A_{10} 
    - A_{20} (\hat{u} \cdot \hat{x}_b) s 
    - A_{11} v_0 \text{e}^{-B_{01} t} \right).  \label{eq:dudt}
\end{equation}
Thus, the required transplant size rates $\alpha$ and $\beta$ in 
Fig.~\ref{fig5} are approximately
$\alpha =
\frac{\text{d}s(0)}{\text{d}t}\big|_{s = s_c}$, and $\beta=
\frac{\text{d}s(\infty)}{\text{d}t}\big|_{s = s_c}$, where $s_c$ is the
transplant size required at the crossover point between these rates (e.g. as
shown in Fig.~\ref{fig5}).

For an initial condition with $v_0 < 0$, which occurs in Fig.~\ref{fig1} when a
nearly healthy state is depleted by antibiotics, $\alpha >
\beta$. In this case the required transplant size monotonically increases until
it attains $s^*$ at the infected steady state, so it is best to administer
FMT as soon as possible. Alternatively, when $v_0 > 0$, $\alpha <
\beta$. When $A_{11} v_0$ is sufficiently large $\alpha$ becomes negative,
which indicates there is a nonzero transplant time at which the required
transplant size is minimized (corresponding to $\frac{\text{d}s}{\text{d}t} =
0$). The concave-up trajectories in Fig.~\ref{fig5} exhibit this optimal
transplant time. For $v_0 > 0$ and under the same conditions for which
Eq.~(\ref{eq:dudt}) was derived, this optimal transplant time
$t^*$ is
\begin{equation}
    t^*_{\text{opt}} = \frac{1}{B_{01}} \ln \left( \frac{A_{11} v_0}{A_{10} - A_{20}
    u_0} \right).
    \label{eq:tstar}
\end{equation}
This nonzero transplant time reflects ecological pressures that temporarily
drive the system closer to the separatrix, overpowering the slow unstable
manifold.  Two trajectories that numerically
recapitulate these two cases are shown in Fig.~\ref{fig3}.

\section{SSR applied to fecal microbiota transplantation} \label{11d_transient}

\begin{figure}[t]
    \includegraphics[width=.47\textwidth]{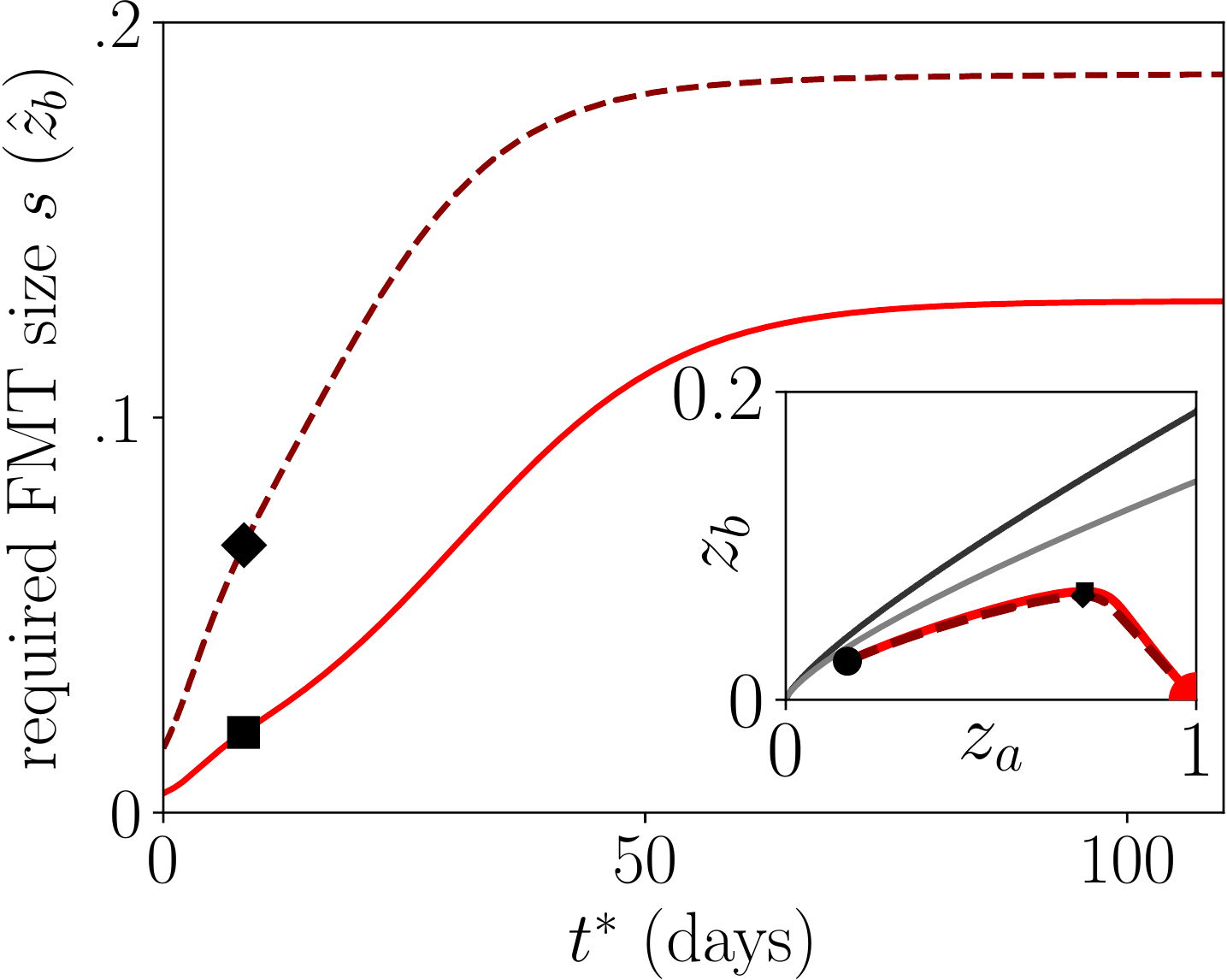}%
    \caption{Transient dynamics are preserved under SSR. (inset)
    Microbial trajectories of the CDI model (in-plane projection,
    dashed) and the associated SSR-reduced model (solid) as in Fig.~\ref{fig4}
    are shown.
    (main panel) At each time along these trajectories, the minimum FMT
    size required to make the state health-prone is plotted, for a transplant made up of
    $\vec{y}_b$ (11D, dashed) or $(0,\ 1)$ (2D, solid). Phase space is linked to
    the FMT size dynamics by indicating the time at which $z_b$ begins
    to decrease with a solid square (2D) or diamond (11D) in both the inset and
    main figure. Since the required FMT size $s$ is the distance between a
    state and the separatrix, the similarity between the two time courses of
    $s$ indicates that SSR preserves transient dynamics. 
        \label{fig7}}
\end{figure}

\begin{figure}[t]
\begin{center}
    \includegraphics[width=.47\textwidth]{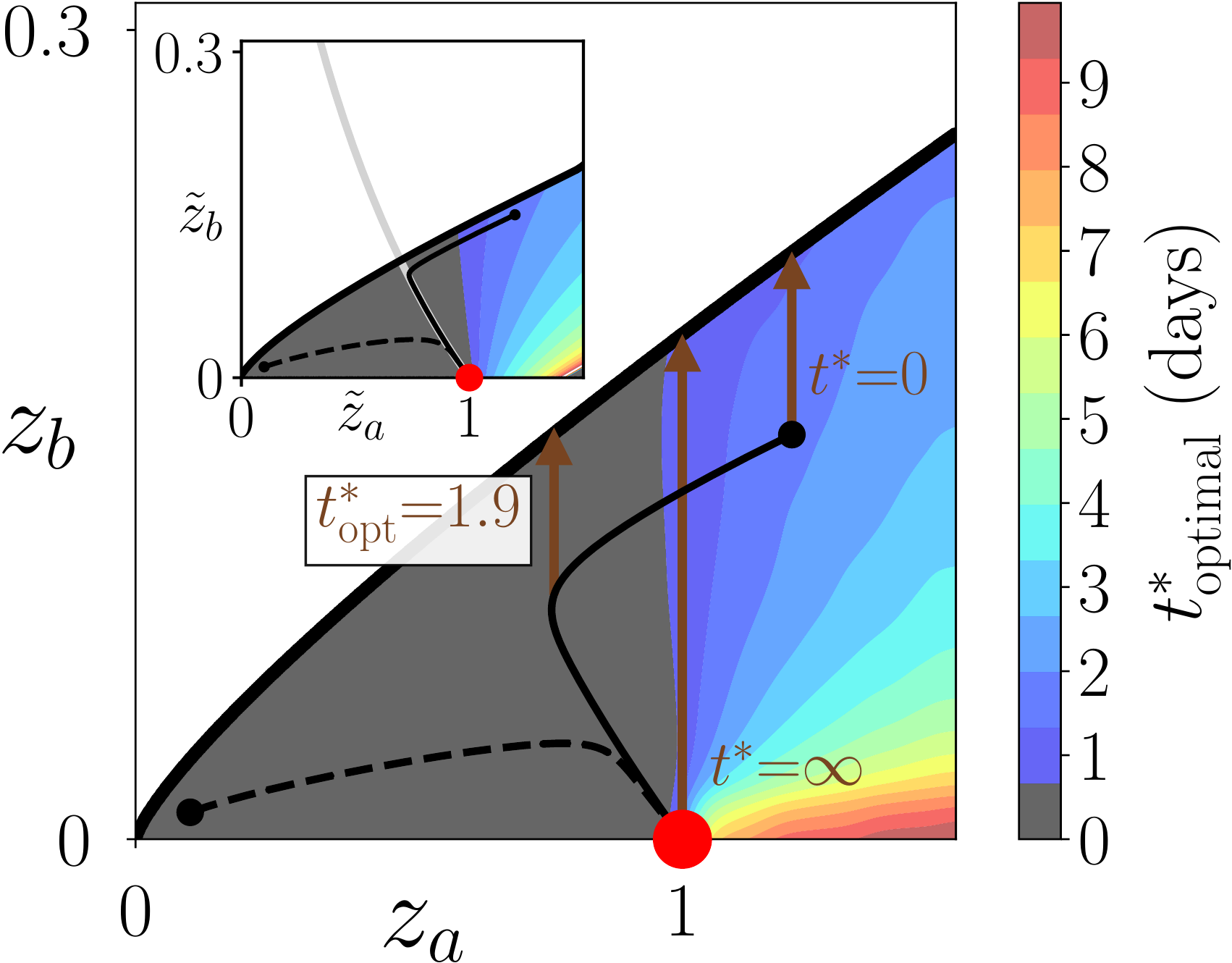}%
    \caption{\edit{Optimal transplant times are preserved under SSR. Optimal
    transplant times $t^*_{\text{opt}}$ of the 11D Stein model (main panel)
    largely match the predictions of its associated SSR-reduced model (inset).
    In the high-dimensional model, $t^*_{\text{opt}}$ is computed numerically
    (as in Fig.~\ref{fig7}) for a grid of points on the plane spanned by $\vec{y}_a$ and
    $\vec{y}_b$ for disease-prone initial conditions (located underneath the
    separatrix, which is shown as a thick black line). The spatial and temporal
    resolutions of this simulation are $\delta z_a = 0.025$, $\delta z_b =
    0.01$, and $\delta t_{\text{opt}} = 0.15$, and the resulting data points
    were smoothed with a Gaussian filter. (inset) We display the optimal
    transplant times of the corresponding SSR-reduced model, as in
    Fig.~\ref{fig3}. The
    SSR-reduced parameters were nondimensionalized so that $t^*_{\text{opt}}$
    could be generated with Eq.~(S88), and the resulting optimal transplant time
    predictions were redimensionalized and plotted. The inset and the main
    panel share the same colorbar.} \label{fig:11D_opt}}
\end{center}
\end{figure}

In Section \ref{sec:2d}, FMT restored a CDI-prone
microbial state in a 2D gLV model. In previous work \cite{JonesCarlson2018}, we
implemented FMT in the previously mentioned 11D CDI model
\cite{SteinXavier2013} and observed similar success. Here, the
behavior of FMT in the CDI model and in its SSR counterpart are shown to match closely,
which indicates that SSR preserves transient microbial dynamics.

Fig.~\ref{fig7} (inset) contains the in-plane projections of the 11D (dashed)
and corresponding SSR-reduced 2D (solid) microbial trajectories with initial
conditions that lie on the plane spanned by $(\hat{y}_a,\ \hat{y}_b)$ (11D) and 
$(\hat{z}_a,\ \hat{z}_b)$ (2D), as in Fig.~\ref{fig4}.
Fig.~\ref{fig7} (main
panel) plots the
required transplant size $s$ at each state along the two
trajectories: the 11D (dashed) transplant 
is composed of $\hat{y}_a$, and $s$ is calculated
numerically with a bisection method; the 2D (solid) transplant is composed of
$\hat{z}_b = (0,\ 1)$, and $s$ is computed analytically with 
Eq.~(\ref{eq:sep}).

In both systems, the microbial trajectories follow a fast stable manifold before
switching to a slow manifold of some hyperbolic fixed point. As in the
2D case, the flow along these fast and slow manifolds underpins
how the required transplant size $s$ changes over time. In Fig.~6, the transition
between the fast and slow manifolds occurs at 8.37 days in 11D (solid
diamond, main panel and inset) and at 8.31 days in 2D (solid square).

\edit{As in the 2D case, the transition between these manifolds may result in
a nonzero optimal transplant time $t^*_{\text{opt}}$.  The main panel of
Fig.~\ref{fig:11D_opt}
displays these optimal transplant times over a range of initial conditions,
in which $t^*_{\text{opt}}$ is generated with the same numerical bisection method as
previously mentioned. Many of the high-dimensional initial conditions exhibit a
non-zero optimal
transplant time, mirroring the results of Fig.~\ref{fig3}. Further, the high-dimensional optimal
transplant times closely match those predicted by SSR, which are displayed in the
inset of Fig.~\ref{fig:11D_opt}, and which were analytically calculated with Eq.~(S88).
}

\edit{Since the SSR-reduced system largely preserves the high-dimensional
transplant time dynamics, and since in the 2D nondimensionalized system
$t^*_{\text{opt}}$ can be examined analytically, the high-dimensional optimal transplant
times may be approximated in terms of the high-dimensional interaction parameters.
First, for systems well-approxiated by SSR, a nonzero optimal transplant time
can only exist when $v_0 > 0$--- this tends to occur when the size of the
initial condition is larger than that of the steady state $\vec{y}_b$, and when its
composition is similar to that of $\vec{y}_b$. For this class of initial
conditions,
$t_{\text{opt}}$ will be smaller when the eigenvalues of the semistable fixed
point ($A_{10}$ and $B_{01}$) are larger, or in terms of the SSR-reduced
parameters, when $M_{ab}/M_{bb}$ and $M_{ba}/M_{aa}$ are larger. }

\edit{The similarities between the transient dynamics of the high-dimensional
and 2D systems, as well as the correspondence in optimal transplant timings,
suggest} that the theoretical analyses of Section IV may inform more
complex and highly-resolved systems.

\section{Discussion}
\subsection{Compression of complex ecological systems}
SSR differs from other model reduction techniques \cite{Gu2011,
GoekeZerz2017} since it preserves core observable ecological features of the
original model, namely steady states and their stabilities.  The behavior of
the model on the transition between two of these steady states is approximated
by SSR. \edit{Though the implementations of SSR demonstrated in
this paper were accurate, in general the accuracy of SSR is not obvious
\textit{a priori}; therefore, in future work it is important to carefully examine the
circumstances under which SSR is effective. When SSR is
accurate,} properties of the steady states the original model may also be
extracted from this approximation--- for example, the size of the basins of
attraction in the approximate system can inform the robustness of a given state
in the original system, and the separatrix of the reduced model can approximate
the slow manifold on which dynamics evolve in the original model. The speed-up
gained by leveraging the analytic tractability of these approximate systems
highlights the utility of SSR.

Beyond applications to existing gLV models, SSR-based methods could
create two-state gLV systems from raw microbial data by choosing basis vectors
during the fitting process that correspond to experimentally observed steady
states \cite{CosteaBork2018}. The resulting models would describe interactions
between steady states rather than between individual species, and would
consist of fewer variables and parameters that have improved explanatory power.
This perspective--- which effectively changes the basis vectors of a gLV model
from species to steady states--- may inform the transitions between
steady states in ecological models.

\subsection{Simplification of gLV-based FMT frameworks}
Bacteriotherapy is a promising frontier of medicine that relies on the notion
that the microbiome's composition can both influence and be influenced by
disease. Then, the deliberate alteration of a dysbiotic microbiome, by FMT for
example, might be a viable treatment option for a range of diseases
\cite{Young2017, BelizarioNapolitano2015}.  Since FMT does not contribute to
antimicrobial resistance, it is an emerging alternative to
antibiotics \cite{BorodyKhoruts2011, HeathTahan2018}.  Clinical studies
continue to regularly identify new diseases that are treated by FMT
\cite{HudsonLamb2017,BilinskiGrzegorz2017,TaurXavier2018}.

In this paper we examined a bistable two-state gLV model from a clinical
perspective, in which interventions such as FMT or antibiotics altered the
outcome of a microbial trajectory.  The
tractability of this two-state system allowed for an explicit understanding of
how the efficacy of FMT is
influenced by the timing of its
administration following antibiotic
treatment. In this model, delaying the administration of FMT in
disease-prone microbiomes could lead to its failure.  Modifying the time course
of a treatment has innovated treatment strategies in cancer immunotherapy
\cite{MessenheimerFox2017} and HIV vaccination \cite{Wang2017}, and the results
of this two-state ecological model suggest that treatment timing may be
relevant for bacteriotherapy as well.

Indeed, some circumstantial evidence exists that supports the predictions of
the two-state model, in which FMT efficacy is improved when administered
shortly after antibiotics.  Kang et al. \cite{KangKrajmalnik2017} used a
promising variant of FMT to induce seemingly long-term changes in the
microbiome and symptoms of children with autism spectrum disorders. This FMT
variant ``Microbiota Transfer Therapy'' first prescribed
the antibiotic vancomycin for two weeks, then bowel cleaning, then a large FMT
dose of Standardized Human Gut Microbiota, and finally two months of daily
maitenance FMT doses. In their study, they intended for the efficacy of FMT to
be improved by first clearing out the microbiome with antibiotics, which is
consistent with the results of the 2D gLV system.  However, future
experiments are needed to quantitatively test the extent to which
antibiotic-depleted states are receptive to FMT-like therapies.

\section{Conclusion}
Broadly, SSR realizes a progression towards the simplification of
dynamical systems: while linearization approximates a dynamical system about a
single steady state, SSR approximates a dynamical system about \textit{two}
steady states. We have shown that SSR produces the best possible in-plane 2D
gLV approximation to high-dimensional gLV dynamics. Further, we have
demonstrated the extent to which the 2D model captures the basins of attraction
and transient dynamics of an experimentally derived model. In addition to
the computational efficiency of this technique, which employs analytic results
rather than expensive simulations, SSR builds an intuition for the
high-dimensional system out of connected 2D cross-sections. 

By \edit{approximating} this complex and classic ecological model with 
analytically tractable ecological subspaces, SSR anchors a high-dimensional
system to well-characterized 2D systems. 
Consequently, this technique offers to unravel the complicated
landscapes that accompany complex systems and their behaviors.

\begin{acknowledgments}
This material was based upon work supported by the National Science Foundation
    Graduate Research Fellowship Program under Grant No. 1650114. Any opinions,
    findings, and conclusions or recommendations expressed in this material are
    those of the author(s) and do not necessarily reflect the views of the
    National Science Foundation. This work was also supported by the David and
    Lucile Packard Foundation and the Institute for Collaborative
    Biotechnologies through contract no. W911NF-09-D-0001 from the U.S. Army
    Research Office. The funders had no role in study design, data collection
and analysis, decision to publish, or preparation of the manuscript.
\end{acknowledgments}

\appendix*

\section{Derivation of Steady State Reduction}
Consider an N-dimensional gLV system given by Eq.~(\ref{eq:gLV_N}) that
exhibits steady states $\vec{y}_a$ and $\vec{y}_b$, with dynamics
given by $\frac{\text{d}\vec{y}}{\text{d}t} = \sum_{i=1}^N
\frac{\text{d}y_i}{\text{d}t} \hat{y}_i$. As in the main text, define
variables $x_a$ and $x_b$ in the direction
of the unit vectors $\hat{x}_a \equiv \vec{y}_a/\lVert \vec{y}_a \rVert_2$, and $\hat{x}_b
\equiv \vec{y}_b/\lVert \vec{y}_b \rVert_2$, where $\lVert \cdot \rVert_k$ is the
$k$-norm. Further consider
the \textit{in-plane} 2D gLV dynamics that exist on the plane spanned by
$\hat{x}_a$ and $\hat{x}_b$. Here, we prove
that the parameters prescribed by Steady State Reduction, given in
Eq.~(\ref{eq:SSR_params}), minimize the 2-norm of the deviation $\vec\epsilon$
between the high-dimensional and in-plane dynamics at every point on the plane.

Consider coefficients $\textbf{c} = \{c_1,\
\ldots,\ c_6\}$ that parameterize the 2D gLV equations,
\begin{equation}
    \begin{split}
        \frac{\text{d}x_a}{\text{d}t} &= x_a\left( c_1 + c_2 x_a + c_3 x_b
        \right), \quad \text{and}  \\
        \frac{\text{d}x_b}{\text{d}t} &= x_b\left( c_4 + c_5 x_a + c_6
        x_b\right), 
    \end{split}
\end{equation}
so that the in-plane dynamics are $\frac{\text{d}\vec{x}}{\text{d}t} =
\frac{\text{d}x_a}{\text{d}t} \hat{x}_a + \frac{\text{d}x_b}{\text{d}t}
\hat{x}_b$. Any point on this plane can be written $\vec{y}
= \vec{y}_a x_a + \vec{y}_b x_b$. 

The deviation between the high-dimensional and in-plane dynamics
$\vec\epsilon$ is
\begin{equation}
    \vec\epsilon(x_a,\ x_b) = \frac{\text{d}\vec{x}}{\text{d}t} -
    \frac{\text{d}\vec{y}}{\text{d}t},
\end{equation}
which is defined at every point on the plane $(x_a,\ x_b)$.
We will show that the parameters precribed by SSR minimize the 2-norm of this
deviation $\lVert \vec\epsilon \rVert_2$ at point on the plane.

The deviation $\vec\epsilon$ can be decomposed into the N-dimensional unit
vectors $\hat{y}_i$, so that $\vec\epsilon = \sum_i \hat{y}_i \epsilon_i$,
where the components $\epsilon_i$ are given by
\begin{widetext}
\begin{equation}
    \begin{split}
        \epsilon_i &=
        y_{ai}x_a\left( (c_1 - \rho_i) + \left(c_2 - \sum_{j=1}^{N}
        K_{ij}y_{aj}\right) x_a + \left(c_3 - \sum_{j=1}^N K_{ij} y_{bj}
        \right) x_b \right) \\
        & \quad \quad \quad \ +
        y_{bi} x_b \left( (c_4 - \rho_i) + \left(c_5 - \sum_{j=1}^{N}
        K_{ij}y_{aj} \right)x_a  + \left(c_6 - \sum_{j=1}^N K_{ij} y_{bj}
        \right)x_b \right) \\
        &\equiv \epsilon_{10,i} x_a + \epsilon_{20,i} x_a^2 +
        \epsilon_{11,i}
        x_a x_b + \epsilon_{01,i}
        x_b + \epsilon_{02,i} x_b^2,
    \end{split}
\end{equation}
\end{widetext}
where components $\epsilon_{jk,i}$ are defined to correspond to contributions
by $x_a^j x_b^k$ terms. Here, $y_{ai}$ corresponds to the $i$th
component of the unit vector $\hat{x}_a \equiv \vec{y}_a / \lVert \vec{y}_a
\rVert_2$.
In the same way, the deviation vector may be decomposed
according to
\begin{equation}
    \vec\epsilon = \vec\epsilon_{10} x_a + \vec\epsilon_{20} x_a^2 +
\vec\epsilon_{11} x_a x_b + \vec\epsilon_{01} x_b + \vec\epsilon_{02} x_b^2.
\end{equation}

Minimizing this deviation at each point $(x_a,\ x_b)$ is equivalent to
minimizing each orthogonal contribution $\vec\epsilon_{jk}$.  Each contribution
is a function of one or two parameters ($\vec\epsilon_{10}(c_1)$,
$\vec\epsilon_{20}(c_2)$, $\vec\epsilon_{01}(c_4)$, $\vec\epsilon_{02}(c_6)$,
and $\vec\epsilon_{11}(c_3,\ c_5)$), which simplifies the minimization process. 

We now find the set of optimal coefficients
$\textbf{c}^* = \{c_1^*,\ \ldots,\ c_6^*\}$ that minimize the 2-norm of each
contribution $\lVert \vec\epsilon_{jk} \rVert_2$. For convenience,
we equivalently minimize the square of this 2-norm. 
The Hadamard square represents the element-wise square of a
vector, defined as $\vec{v}^{\ \circ 2} = [v_1^2,\ v_2^2,\ \ldots,\
v_N^2]^T$.

The coefficient $\lVert \vec\epsilon_{10} \rVert_2^2$ is given by
\begin{equation}
    \lVert \vec\epsilon_{10} \rVert_2^2 = \sum_{i=1}^N y_{ai}^2 (c_1 -
    \rho_i)^2.
\end{equation}
When minimized with respect to $c_1$, this quantity obeys 
\begin{equation}
    \frac{\text{d}\lVert \vec\epsilon_{10} \rVert_2^2}{\text{d}c_1} =
    \sum_{i=1}^N 2 y_{ai}^2 (c_1 -
    \rho_i) = 0,
\end{equation}
which is satified for
\begin{equation}
    c_1^* = \frac{ \sum_{i=1}^N y_{ai}^2 \rho_i}{\sum_{i=1}^N y_{ai}^2}
    = \frac{ \vec{y}_{a}^{\ \circ 2} \cdot \vec\rho}{\lVert \vec{y}_a
    \rVert_2^2}.
\end{equation}
In a similar way, $\lVert \vec\epsilon_{20} \rVert_2^2$, $\lVert
\vec\epsilon_{01} \rVert_2^2$, and $\lVert \vec\epsilon_{02} \rVert_2^2$ are
minimized when
\begin{equation}
    \begin{split}
    c_2^* &= \frac{ \sum_{i=1}^N \left( y_{ai}^2 \sum_{j=1}^N
    K_{ij}y_{aj} \right)}{\sum_{i=1}^N y_{ai}^2}
        = \frac{ (\vec{y}_{a}^{\ \circ 2})^T K \vec{y}_a}{\lVert \vec{y}_a
    \rVert_2^3},
    \end{split}
\end{equation}
\begin{equation}
    \begin{split}
    c_4^* = \frac{ \sum_{i=1}^N y_{bi}^2 \rho_i}{\sum_{i=1}^N y_{bi}^2}
    = \frac{ \vec{y}_{b}^{\ \circ 2} \cdot \vec\rho}{\lVert \vec{y}_b
    \rVert_2^2},
    \end{split}
\end{equation}
and
\begin{equation}
    \begin{split}
    c_6^* &= \frac{ \sum_{i=1}^N \left( y_{bi}^2 \sum_{j=1}^N
    K_{ij}y_{bj} \right)}{\sum_{i=1}^N y_{bi}^2}
        = \frac{ (\vec{y}_{b}^{\ \circ 2})^T K \vec{y}_b}{\lVert \vec{y}_b
    \rVert_2^3}.
    \end{split}
\end{equation}

Lastly, the squared norm of the cross-term $\lVert
\vec\epsilon_{11}\rVert_2$ is given by
\begin{equation}
    \begin{split}
        \lVert \vec\epsilon_{11} \rVert_2^2 &= 
    \sum_{i=1}^N \left[ y_{ai} \left(c_3 - \sum_{j=1}^N K_{ij} y_{bj}
    \right) \right. \\ 
        &\quad \quad \ \quad \left. 
        + y_{bi} \left(c_5 - \sum_{j=1}^N K_{ij} y_{aj}
        \right) \right]^2.
    \end{split}
\end{equation}
Minimizing with respect to $c_3$ and $c_5$ results in
\begin{equation}
    \begin{split}
        \frac{\text{d} \lVert \vec\epsilon_{11} \rVert_2^2}{\text{d}c_3} &= 
    \sum_{i=1}^N 2 \left[ y_{ai}^2 \left(c_3 - \sum_{j=1}^N K_{ij} y_{bj}
    \right) \right. \\ 
        &\quad \quad \ \ \quad \left. 
        + y_{ai} y_{bi} \left(c_5 - \sum_{j=1}^N K_{ij} y_{aj}
        \right) \right] \\
        &= 0,
    \end{split}
\end{equation}
and
\begin{equation}
    \begin{split}
        \frac{\text{d} \lVert \vec\epsilon_{11} \rVert_2^2}{\text{d}c_5} &= 
        \sum_{i=1}^N 2 \left[ y_{ai}y_{bi} \left(c_3 - \sum_{j=1}^N K_{ij} y_{bj}
    \right) \right. \\ 
        &\quad \quad \ \ \quad \left. 
        + y_{bi}^2 \left(c_5 - \sum_{j=1}^N K_{ij} y_{aj}
        \right) \right] \\
        &= 0.
    \end{split}
\end{equation}

After rearranging terms, these conditions read
\begin{equation}
    \begin{split}
        c_3 \sum_{i=1}^N y_{ai}^2 + c_5 \sum_{i=1}^N y_{ai} y_{bi} &= \sum_{i=1}^N 
        \sum_{j=1}^N y_{ai} K_{ij} (y_{ai} y_{bj} + y_{bi} y_{aj}),
    \end{split} \end{equation}
and
\begin{equation}
    \begin{split}
        c_3 \sum_{i=1}^N y_{ai} y_{bi} + c_5 \sum_{i=1}^N y_{bi}^2 &= \sum_{i=1}^N 
        \sum_{j=1}^N y_{bi} K_{ij} (y_{ai} y_{bj} + y_{bi} y_{aj}), \\
    \end{split}
\end{equation}
which are satisfied when
\begin{equation}
    \begin{split}
        c_3^*  &= 
        \frac{\sum_{i,j=1}^N K_{ij} (y_{ai} y_{bj} + y_{bi} y_{aj})
        \left(\sum_{k=1}^N y_{ai} y_{bk}^2 - y_{bi} y_{ak} y_{bk} \right)}
        { \left(\sum_{i=1}^N y_{ai}^2\right) \left(\sum_{i=1}^N y_{bi}^2\right)
        - \left(\sum_{i=1}^N y_{ai} y_{bi} \right)^2 }, \\
    \end{split}
\end{equation}
and
\begin{equation}
    \begin{split}
        c_5^*  &= 
        \frac{\sum_{i,j=1}^N K_{ij} (y_{ai} y_{bj} + y_{bi} y_{aj})
        \left(\sum_{k=1}^N y_{bi} y_{ak}^2 - y_{ai} y_{ak} y_{bk}  \right)}
        { \left(\sum_{i=1}^N y_{ai}^2\right) \left(\sum_{i=1}^N y_{bi}^2\right)
        - \left(\sum_{i=1}^N y_{ai} y_{bi} \right)^2 }. \\
    \end{split}
\end{equation}
However, when $\vec{y}_a$ and $\vec{y}_b$ are orthogonal, the cross-term
deviation $\lVert \vec \epsilon_{11} \rVert_2^2$ is simplified, and the optimal
coefficients $c_3^*$ and $c_5^*$ become
\begin{equation}
    \begin{split}
    c_3^* &= \frac{ \sum_{i=1}^N \left( y_{ai}^2 \sum_{j=1}^N
    K_{ij}y_{bj} \right)}{\sum_{i=1}^N y_{ai}^2}
        = \frac{ (\vec{y}_{a}^{\ \circ 2})^T K \vec{y}_b}{\lVert \vec{y}_a
    \rVert_2^2 \lVert \vec{y}_b
    \rVert_2},
    \end{split}
    \label{eq:c3}
\end{equation}
and
\begin{equation}
    \begin{split}
    c_5^* &= \frac{ \sum_{i=1}^N \left( y_{bi}^2 \sum_{j=1}^N
    K_{ij}y_{aj} \right)}{\sum_{i=1}^N y_{bi}^2}
        = \frac{ (\vec{y}_{b}^{\ \circ 2})^T K \vec{y}_a}{\lVert \vec{y}_b
    \rVert_2^2 \lVert \vec{y}_a
    \rVert_2}.
    \end{split}
    \label{eq:c6}
\end{equation}

Since the squared norms of the deviations $\lVert \epsilon_{jk} \rVert_2$ are
convex, the coefficient set \textbf{c}$^*$ is a global
minimum for $\lVert \vec\epsilon \rVert_2$. Therefore, we have identified the
parameters that minimize the deviation between the in-plane and
high-dimensional gLV dynamics for any point on the plane spanned by $\vec{y}_a$
and $\vec{y}_b$.

\end{document}